\documentclass[aps,pra,twocolumn,showpacs,superscriptaddress]{revtex4}
\usepackage{graphicx}
\usepackage{amsmath}
\usepackage{amssymb}

\input{epsf}
\begin{document}

\title{The Trapped Polarized Fermi Gas at Unitarity}

\author{D. Blume}
\affiliation{Department of Physics and Astronomy,
Washington State University,
  Pullman, Washington 99164-2814, USA}

\date{\today}

\begin{abstract}
We consider  
population-imbalanced two-component Fermi gases under
external harmonic confinement interacting through
short-range two-body potentials with diverging 
$s$-wave scattering length. Using the fixed-node
diffusion Monte Carlo method,
the energies of the ``normal state''
are determined as functions of the population-imbalance and
the number of particles. 
The energies of the trapped system follow, to a good approximation,
a universal curve even for fairly small systems. A simple
parameterization of the
universal curve is presented and related to the equation of state
of the bulk system.
\end{abstract}


\maketitle

Strongly-correlated few- and many-body systems play
an important role
in
atomic, 
nuclear and condensed matter physics
(see, e.g., Refs.~\cite{tann00,gior07,carl98,kotl06}).
These
systems
are characterized by 
intricate particle-particle correlations,
which can 
be difficult to capture by
mean-field frameworks. Furthermore, 
the development of effective beyond mean-field
approaches is 
complicated by the fact that a small parameter can typically not be
identified. 
Thus, the most promising avenues for the theoretical 
description of strongly-correlated systems start with a microscopic many-body
Hamiltonian. In a few fortuitous cases,
such as strongly-correlated one-dimensional systems~\cite{matt93}, exact 
analytical solutions can be found for certain classes of model
Hamiltonian.
In other cases, however, a quantitative description
of strongly-correlated systems relies on numerical approaches.

This paper treats strongly-correlated two-component
Fermi gases under external harmonic confinement with diverging
$s$-wave scattering length $a_s$ at zero temperature
using a numerical Monte Carlo approach. 
The masses of the two species are assumed to be the same,
and the properties of the system are determined
as a function of the population difference between the 
two components.
The interest in population-imbalanced atomic Fermi 
systems~\cite{theory,lobo06,prok08,pila08,reca08,ketterle,hulet}
is inherently linked to
the mismatch of the Fermi surfaces of the two
components, which has, e.g., been predicted to lead 
under certain circumstances to
pairing with non-zero momentum, i.e., to the
formation of so-called FFLO states~\cite{fuld64,lark65}. 
So far, 
the experimental search for FFLO
states 
has, however, not been met with success.

Population-imbalanced Fermi systems have
been
realized using  ultracold atom samples, and a rich behavior
as functions of population-imbalance and temperature has been 
observed~\cite{ketterle,hulet}.
In these experiments, 
composite fermionic atoms such as $^6$Li 
are trapped in two different internal hyperfine states,
referred
to as spin-up and spin-down atoms in the following. The population-imbalance
or polarization 
can be adjusted straightforwardly, and
the interspecies
scattering length $a_s$ between the spin-up and spin-down fermions can
be varied by applying an external magnetic field in the vicinity
of a so-called Fano-Feshbach resonance. To date, all experiments 
on population-imbalanced Fermi gases have been
performed with comparatively large atom samples. Here, we consider 
the properties of small
systems, which can be realized using present-day technology
by loading a degenerate 
Fermi gas into a deep optical lattice, for which the tunneling between
neighboring sites is neligible.

We determine the 
energetics
of small trapped two-component Fermi gases at unitarity
with varying population-imbalance
using the fixed-node diffusion Monte Carlo (FN-DMC) method~\cite{hamm94}.
Two different parameterizations of the many-body nodal 
surface 
are considered:
The nodal surface 
of the non-interacting 
trapped gas, and a nodal surface that is constructed by 
anti-symmetrizing a set of pair functions.
Not surprisingly, the trapped 
gas described by the former nodal surface 
can be related to the normal state of the homogeneous 
system.
We find that the energies of the ``trapped normal gas''
with varying population-imbalance and number of particles
fall on a universal curve that can be parameterized quite well
by three parameters. 
Remarkedly, the energies
for systems with as few as $N=5$ fermions fall
on the same universal curve 
as those
for larger systems (the largest $N$ considered 
in this work is 55).
The relationship between the trapped and homogeneous
systems is also analyzed by considering structural 
properties.
For small $N$, 
we find that the energies of the trapped normal
system are lower than those obtained for a nodal
surface 
that accounts for pairing physics.
For larger $N$, in contrast,
the nodal surface that accounts for pairing 
physics results in a lower energy for small $|N_1-N_2|$, 
where
$N_1$ and $N_2$ denote the number of spin-up
and spin-down fermions, respectively. 
Our {\em{ab initio}} results 
for trapped population-imbalanced Fermi gases
may
serve as benchmarks 
for other numerically less involved techniques
and
aid in testing phenomenological models.

Our model Hamiltonian for the trapped two-component Fermi gas with 
$N$ mass $m$ fermions, where $N=N_1+N_2$ and
$N_1 \ge N_2$, reads
\begin{eqnarray}
\label{eq_ham}
H = 
\sum_{i=1}^{N} \left(
\frac{-\hbar^2}{2m} \nabla^2_{\vec{r}_i} + \frac{m \omega^2}{2} 
\vec{r}_i^2 \right) + 
\sum_{i=1}^{N_1} \sum_{j=N_1+1}^{N} V_{tb}({r}_{ij}).
\end{eqnarray}
Here,
$\vec{r}_i$ denotes the position
vector of the $i$th atom and $\omega$ the
angular trapping frequency.
The interaction potential $V_{tb}$ between spin-up and
spin-down atoms depends on the interparticle
distance ${r}_{ij}$, ${r}_{ij} =  |\vec{r}_i - \vec{r}_{j}|$,
and is characterized by 
the interspecies $s$-wave scattering length $a_s$.
We model $V_{tb}$ 
by a square well potential with range $R_0$ and depth $V_0$ ($V_0>0$),
$V_{tb}(r) = -V_0 \mbox{ for } r < R_0$
and
$0 \mbox{ for } r > R_0$.
To describe the unitary regime, we fix $R_0$ at
$R_0 = 0.01 a_{ho}$, where $a_{ho}$ denotes
the harmonic
oscillator length, $a_{ho}=\sqrt{\hbar/(m \omega)}$, 
and adjust the depth $V_0$ 
so that the free-space two-particle system supports a single
zero-energy $s$-wave bound state. 
We consider regimes away from an intra-species
$p$-wave Feshbach resonance, and treat
like atoms as
non-interacting.

The ground state wave function for fermions is,
as a consequence
of the Pauli exclusion principle,
 characterized by
a complicated nodal surface, i.e.,
the many-body wave function changes sign when
either two spin-up atoms or two spin-down atoms are exchanged.
This sign change often times leads to 
inefficient 
sampling schemes,
a
phenomenon  
commonly referred to as ``fermionic sign problem''.
To avoid  the sign problem, we adopt the FN-DMC method~\cite{hamm94},
which determines the eigenenergy of a state that has
the same nodal structure as $\psi_T$ but that may
differ from $\psi_T$ in other regions of
configuration space. 
In most applications of the FN-DMC method,
great effort is placed on optimizing $\psi_T$ so as to
obtain a tighter upper bound to the true
eigenenergy.
Our calculations that utilize a pairing function
fall into this category.
Our study of the trapped normal state,
in contrast,
considers the properties of the many-body
system for a fixed, non-optimized nodal structure.

The guiding function $\psi_{T1}$ of 
the trapped normal gas 
is written in terms of the
ground state wave function $\psi_{NI}$ of
the 
non-interacting 
trapped 
Fermi 
gas 
and three Jastrow factors
$J_{kk'}$,
$\psi_{T1} = \psi_{NI} J_{11} J_{22} J_{12}$.
The positive definite Jastrow factors $J_{kk'}$ account for 
correlations between atoms from component $k$ and atoms 
from component $k'$;
they 
reduce the statistical uncertainties but
do not alter the nodal surface of 
$\psi_{T1}$. The function
$\psi_{NI}$ can be written as a product of
two Slater determinants,
$\psi_{NI} =\mbox{Det}(M_1) \times \mbox{Det}(M_2)$,
where the $ij$th element of 
the matrix $M_k$, $k=1$ or 2,
is given by the single particle harmonic oscillator function
$\phi_i(\vec{r}_j)$
with $i,j=1,\cdots,N_k$.
The subscript $i$
labels the excitations: $i=1$ corresponds to the
ground state, $i=2$ to the first excited state, and so on.
For closed shells, i.e., for $N_k=1,4,10,20,35,\cdots$,
the Slater determinant $\mbox{Det}(M_k)$ is uniquely defined. For open shells,
in contrast, degenerate states exist and
$\psi_{NI}$
is not uniquely defined. 
In this case,
we consider determinants build from
different sets of harmonic oscillator orbitals and report the 
lowest energy.
For $N_k=5$, e.g., we fill the first two shells and place the remaining
particle in either a $l=0$ or $l=2$ orbital ($l$ denotes the orbital
angular momentum of $\phi_i$).

Tables~\ref{tab_ennormal} 
\begin{table}
\caption{
FN-DMC energies in units of $\hbar \omega$
for two-component unitary Fermi gases 
with various $N_1,N_2$ combinations
($N \le 20$
and $|N_1-N_2| \le 10$) 
calculated using 
$\psi_{T1}$.
The 
energies are uncertain in the last digit
reported.
The energy of the $(3,1)$ system is
$6.60 \hbar \omega$;
the energies of the other small systems
($N_1, N_2 \le 3$) 
are
reported in Refs.~\cite{blum07,stec08}.
}
\begin{ruledtabular}
\begin{tabular}{l|cccccccccc}
   & 1 & 2 & 3 & 4 & 5 & 6 & 7 & 8 & 9 & 10\\
\hline
4 & 8.93 & 10.2 & 11.4 & 12.6 \\
5 & 12.1 & 13.3 & 14.5 & 15.7 & 17.8 \\
6 & 15.8 & 16.6 & 17.7 & 18.8 & 20.9 & 23.1\\
7 & 19.0 & 19.9 & 20.8 & 21.9 & 23.9 & 26.0 & 28.7 \\
8 & 22.5 & 23.3 & 24.1 & 24.9 & 26.9 & 28.9 & 31.0 & 33.2 \\
9 & 25.9 & 26.6 & 27.2 & 28.1 & 30.0 & 32.0 & 34.0 & 36.1 & 38.2\\
10& 29.4 & 29.9 & 30.5 & 31.2 & 33.1 & 35.0 & 37.1 & 39.1 & 41.2 & 43.2\\ 
\end{tabular}
\end{ruledtabular}
\label{tab_ennormal}
\end{table}
and \ref{tab_ennormal2} summarize our FN-DMC energies for 
\begin{table}
\caption{
FN-DMC energies in units of $\hbar \omega$
for unitary Fermi gases 
with closed shells and $N>20$,
calculated using 
$\psi_{T1}$.
The 
energies are uncertain in the last digit
reported.
}
\begin{ruledtabular}
\begin{tabular}{l|cccc}
   & 1 & 4 & 10 & 20  \\
\hline
20 & 73.78 & 73.79 & 81.7 & 109.7 \\
35 & 155.7 & 154.0 & 158.2 & 178.6 \\
\end{tabular}
\end{ruledtabular}
\label{tab_ennormal2}
\end{table}
the guiding function $\psi_{T1}$.
In addition to the FN-DMC energy,
we calculate the expectation value $E_{VMC}$,
$E_{VMC}= \langle \psi_T | H| \psi_T \rangle / \langle \psi_T | \psi_T 
\rangle$, 
of the
many-body Hamiltonian 
using the variational Monte Carlo (VMC) 
method~\cite{hamm94}.
For all systems considered in Table~\ref{tab_ennormal}
[as well as for the $(N_1,N_2)=(20,1)$, $(35,1)$, $(20,4)$, 
and $(35,4)$ systems],
$E_{VMC}$ is, for the simulation times employed, 
positive and less than about 15~\%
higher than the corresponding FN-DMC energy.
For the systems with 
$(N_1,N_2)=(20,10)$, $(20,20)$, $(35,10)$ and $(35,20)$,
in contrast, the VMC energy becomes negative
after a fairly small number of sampling steps despite the fact that
the initial configurations correspond to a gas-like system. 
To obtain the FN-DMC energies
for these 
systems,
we start the FN-DMC calculations from 
non-equilibrated gas-like configurations 
and not, as done for the smaller systems, from configurations that
are distributed according to $\psi_{T1}^2$.
The resulting FN-DMC energies correspond to a gas-like state
and appear converged.
For the $(20,10)$ system,
e.g., we have checked that different initial configurations result, 
within errorbars, in the same FN-DMC energy.
The
existence of many-body bound states with negative energy 
for large systems interacting
through finite-range two-body potentials is not 
surprising (see, e.g., Ref.~\cite{chan04}).
In fact, even some of the smaller systems
with $N \ge N_{cr}$
may
possess a
tightly-bound molecular-like 
ground state that is not sampled
in our simulations.
For systems 
with $N_1=N_2$, it has been found
previously that
the critical number $N_{cr}$ is larger than 6~\cite{blum07};
for systems with
$N_1-N_2=1$, it is larger than 5~\cite{blum07}.

\begin{figure}
\includegraphics[angle=270,width=80mm]{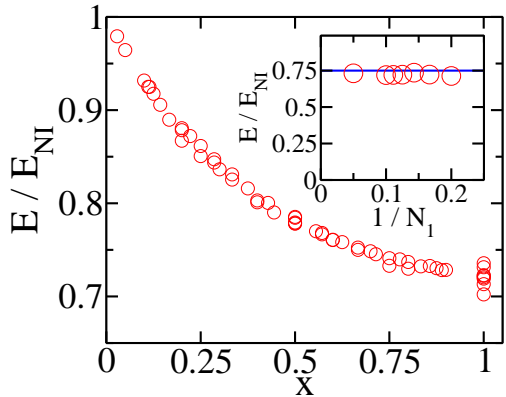}
\caption{
(Color online)
The
scaled energies $E/E_{NI}$
for the trapped normal 
Fermi gas 
at unitarity, shown as a function
of the concentration $x$,
fall to a good approximation on a universal curve.
The inset shows $E/E_{NI}$ 
for $x=1$ as a function of $1/N_1$, $N_1=5-20$;
the LDA prediction $E/E_{NI}=0.75$ is shown by a solid line.} 
\label{fig_ennormal}
\end{figure}

Figure~\ref{fig_ennormal} shows the FN-DMC energies $E$ 
from
Tables~\ref{tab_ennormal} and \ref{tab_ennormal2}, 
scaled by the 
total energy $E_{NI}$ of the corresponding
non-interacting 
gas, as a function of the concentration $x$,
where $x=N_2/N_1$. A fully polarized, single-component
Fermi gas corresponds to $x=0$
while a Fermi gas with equal number of particles
in the two components corresponds to $x=1$.
Notably, the scaled energies $E/E_{NI}$ 
fall to a 
good approximation on a universal curve.
Applying the local density approximation
(LDA) to the equation of state
of the population-balanced homogeneous normal system at unitarity,
the energy of the trapped unitary system with $x=1$ is given by
$\sqrt{\xi}E_{NI}$, where $\xi=0.54$~\cite{carl03} 
[a more recent study reports $\xi=0.56(1)$~\cite{lobo06},
implying $\sqrt{\xi}=0.75(2)$].
The inset of Fig.~\ref{fig_ennormal} shows
the scaled energies $E/E_{NI}$ for 
$x=1$ at unitarity 
as a function of $1/N_1$ ($N_1=5-20$).
The fact that $E/E_{NI}$ 
is 
close to $0.75$
suggests
that 
the 
trapped population-balanced unitary gas, described by a wave function
whose nodal surface coincides with that of the 
non-interacting 
gas,
can be described 
quite accurately
by applying the LDA to the normal state of the homogeneous system.

\begin{figure}
\includegraphics[angle=270,width=80mm]{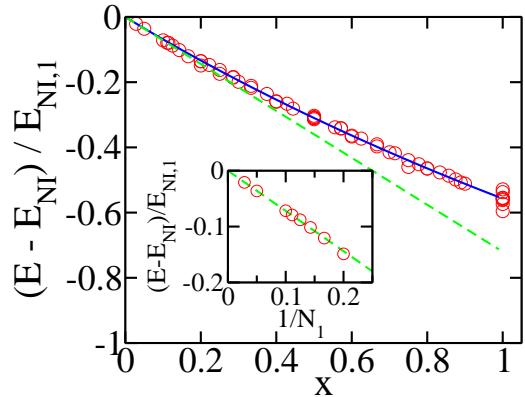}
\caption{
(Color online)
Energetics of the trapped population-imbalanced normal
Fermi gas.
Main figure: Symbols show the quantity $(E-E_{NI})/E_{NI,1}$ as
a function of $x$.
Inset: Symbols show 
$(E-E_{NI})/E_{NI,1}$
for a single down-fermion immersed in a
cloud of up-fermions as a function of 
$1/N_1$ ($N_1=5-35$).
The dashed lines show Eq.~(\ref{eq_fit1}) with
$A_*=0.72$.
The solid line shows Eq.~(\ref{eq_fit2}) with
$A_*=0.72$, $C_*=0.16$ and $D_*=1.7$.
}
\label{fig_impurity}
\end{figure}

To analyze the energies of trapped polarized unitary Fermi gases
further,
we consider the energy of a single down-fermion or impurity 
immersed in a
cloud of up-fermions. Symbols in the inset of
Fig.~\ref{fig_impurity}
show the scaled energy difference $(E-E_{NI})/E_{NI,1}$,
where $E$ is the FN-DMC energy of the $(N_1,1)$ system ($N_1=5-35$),
as a function of $1/N_1$; since $N_2=1$, the quantity $1/N_1$ coincides
with the concentration $x$.
$E_{NI,k}$ denotes the energy
of the 
non-interacting 
single-component 
Fermi gas with $N_k$ atoms ($E_{NI}=E_{NI,1}+E_{NI,2}$).
If the system was 
non-interacting,
the quantity $E-E_{NI}$ would be zero. The deviation from zero can
thus be interpreted as the
attractive interaction or binding energy between the impurity and the
up-fermions.
To determine this energy,
we write
\begin{eqnarray}
\label{eq_fit1}
(E - E_{NI})/E_{NI,1} = -A_* / N_1.
\end{eqnarray}
A linear fit to the FN-DMC energies,
shown by a dashed line in the inset of Fig.~\ref{fig_impurity},
gives $A_*=0.72(1)$.
The analysis performed here for the trapped system is similar to 
that performed for the
homogeneous system~\cite{lobo06,prok08,pila08}. 
However, while we 
fix the impurity mass at its ``bare'' 
value,
work on the homogeneous system
treats the impurity as a quasi-particle with effective
mass $m_{eff}$ and finds $m_{eff}=1.09(2)m$ at 
unitarity~\cite{pila08}.

Symbols in the main
part of Fig.~\ref{fig_impurity} show the 
scaled energy $(E-E_{NI})/E_{NI,1}$
as a function of $x$ (the plot includes
all energies reported in Tables~\ref{tab_ennormal}
and \ref{tab_ennormal2}).
A dashed line shows the quantity
$-A_*x$.
For larger $x$,
the 
scaled
energies lie
above the dashed line, 
indicating a 
shielding of 
the attractive up-down interaction, 
or equivalently, the presence
of an effective repulsion between the down-fermions.
A simple expression for the energy of the normal 
state of trapped two-component systems
with arbitrary $x$ ($x \ge 1/N_1$)
reads
\begin{eqnarray}
\label{eq_fit2}
(E -E_{NI})/E_{NI,1} =  -A_* x + C_* x^{D_*}.
\end{eqnarray}
Using $A_*=0.72$,
a fit to our data
gives $C_*=0.16(1)$ and $D_*=1.7(1)$.
This fit
(solid line in Fig.~\ref{fig_impurity}) 
provides a good description of our numerically
determined energies.
Note that our $D_*$ value would be somewhat larger if
we enforced Eq.~(\ref{eq_fit2}) to reproduce the
LDA value of $0.75$ at $x=1$.

In addition to the energies, we analyze the structural
properties of population-imbalanced Fermi gases.
The LDA predicts to leading order
in $x$ that the
density of the up-fermions is 
unchanged while the down-fermions feel a modified
trapping potential with effective
angular frequency 
$\omega_{eff} = \omega \sqrt{(1+3A_{hom}/5)m/m_{eff}}$~\cite{lobo06,reca08},
where $A_{hom}=0.99(1)$~\cite{pila08}. 
The single down-fermion
immersed in a cloud of up-fermions is thus
predicted to be
described by the ground state 
harmonic oscillator function with 
width $a_{ho,eff}$,
where $a_{ho,eff}=\sqrt{\hbar/(m_{eff} \omega_{eff})}$.
A fit of the square of the 
harmonic oscillator function
to the density profiles of the down-fermion
for the $(10,1)$ and
$(20,1)$ systems,
calculated using the
mixed estimator~\cite{hamm94}, provides
a good description of the density profiles and
gives $a_{ho,eff}=0.88 a_{ho}$ and $0.86a_{ho,eff}$,
respectively, compared with the 
LDA prediction of $a_{ho,eff}=0.87 a_{ho}$.
The good agreement 
suggests that the LDA provides
a valid description even of fairly small
trapped systems.

In addition to the 
guiding function $\psi_{T1}$, we consider the guiding $\psi_{T2}$,
which is constructed by 
anti-symmetrizing a set of
$N_2$  pair functions $\bar{f}$ and $N_1-N_2$
mutually orthogonal 
harmonic oscillator functions $\phi_i$
(recall, $N_1 \ge N_2$)~\cite{stec08,carl03}.
It has been shown previously~\cite{blum07,stec08} that 
$\psi_{T2}$ results in a lower FN-DMC energy  than 
$\psi_{T1}$ for 
population-balanced 
systems with $N \ge 6$
and for odd $N$ systems with $|N_1-N_2|=1$ and $N \ge 13$.
Here, we extend the analysis to systems with larger
$N_1-N_2$.
For the $(7,5)$ system, $\psi_{T2}$ and $\psi_{T1}$ 
result, within errorbars, in the same energy.
For the next larger systems with $N_1-N_2=2$, 
the guiding function $\psi_{T2}$ results in lower FN-DMC energies
than $\psi_{T1}$ [$E= 28.6$, $33.7$, and $38.9 \hbar \omega$
for the $(8,6)$, $(9,7)$ and 
$(10,8)$ systems, respectively].
For $N_1-N_2 \ge 3$ and $N \le 20$, in contrast, we find that the 
guiding function $\psi_{T1}$ 
results in lower FN-DMC energies than $\psi_{T2}$.

In summary, this paper treats 
trapped polarized two-component
Fermi gases 
interacting
through short-range two-body potentials at unitarity. 
This strongly-correlated
regime has attracted a great deal of attention
since the only meaningful length scale
is the system's size
(see, e.g., Ref.~\cite{universal}); 
consequently, the properties of the gas are 
governed by a few universal parameters. Using the
FN-DMC method, we have
determined the energies of the trapped normal gas
as functions of $N$ and $x$.
Our
energies 
fall on a universal
curve that is well described by three parameters.
Guiding functions
that account for pairing physics are also considered
and found to result in lower energies than those
obtained for the trapped normal state 
when $N$ is sufficiently large
and $N_1-N_2$ sufficiently small.
Our results may aid in assessing the 
accuracy of other numerical approaches such as 
density functional theory approaches~\cite{bulg07}. 
Furthermore, they may guide optical lattice experiments
on ultracold fermionic gases.

DB gratefully acknowledges support by the NSF through
grant PHY-0555316.

\end{document}